\documentclass{pspum-l}
\usepackage[cmtip,arrow]{xy}\usepackage{pb-diagram,pb-xy}%
\input xy
\xyoption{all}
\xyoption{all}
 \usepackage{cite}
 \usepackage{graphicx}
 \usepackage{multicol}
 \usepackage{amsfonts}
\usepackage{mathrsfs}
 \usepackage{amssymb}
\usepackage{amsmath}%
\def\cn{\mathcal{N}}
\def\Z{\mathbb{Z}}
\def\C{\mathbb{C}}
\def\cw{\mathcal{W}}
\def\cm{\mathcal{M}}

\begin{document}


\title[The quiver approach to BPS spectra]{The quiver approach to the BPS spectrum\\ of a $4d$ $\mathcal{N}=2$ gauge theory}
\author{Sergio Cecotti}
\address{Scuola Internazionale di Studi Avanzati, via Bonomea 265, I-34100 Trieste, ITALY}
\email{{\tt cecotti@sissa.it}}
\dedicatory{Dedicated to the Memory of Professor Friedrich Hirzebruch}
\date{December, 2012}
\subjclass[2010]{81T60}

\begin{abstract}We present a survey of the computation of the BPS spectrum of a general four--dimensional $\mathcal{N}=2$ 
supersymmetric gauge theory in terms of the Representation Theory of  quivers with superpotential. We focus on SYM with 
a general gauge group $G$ coupled to standard matter in arbitrary representations of $G$ (consistent with a non--positive beta--function).
 The situation is particularly tricky and interesting when the matter consists of an odd number of \emph{half}--hypermultiplets:
 we describe in detail $SU(6)$ SYM coupled to a $\tfrac{1}{2}\,\mathbf{20}$, $SO(12)$ SYM coupled to a $\tfrac{1}{2}\,\mathbf{32}$,
 and $E_7$ SYM coupled to a $\tfrac{1}{2}\,\mathbf{56}$.
\end{abstract}

\maketitle


\section{Introduction}\label{intt}

In the last few years many new  powerful methods were introduced to compute the exact BPS spectrum of a four--dimensional
 $\cn=2$ supersymmetric QFT. We may divide the methods in two broad classes:
 \textit{i)} geometric methods \cite{GMN09,GMN10,GMN11,GMN12,GMN12b} and
\textit{ii)} algebraic methods \cite{CNV,CV11,arnold,ACCERV1,ACCERV2,cattoy,half,nons,nonsX}.
The geometric methods give a deep understanding of the non--perturbative physics, while the algebraic ones are quite 
convenient for actual computations. In the algebraic approach the problem of computing the BPS spectrum is mapped to a \emph{canonical} problem in the Representation Theory (RT) of (basic) associative algebras. A lot of classical results in RT have a direct physical interpretation and may be used to make the BPS spectral problem `easy' for interesting classes of $\cn=2$ theories. Besides, by comparing RT and physics a lot of interesting structures emerge which shed light on both subjects.

\subsection{From $\cn=2$ QFT to quiver representations}\label{qquiq}

To fix the notation, we recall how the BPS states are related to quiver representations, referring to \cite{ACCERV2} for more details. The conserved charges of the theory (electric, magnetic, and flavor) are integrally quantized, and hence take value in a lattice $\Gamma=\oplus_v\, \Z e_v$. On $\Gamma$ we have a skew--symmetric integral pairing, $\langle\gamma, \gamma^\prime\rangle_\text{Dirac}\in \Z$, given by the Dirac electro--magnetic pairing; the flavor charges then correspond to the zero--eigenvectors of the matrix $B_{uv}\equiv \langle e_u, e_v\rangle_\text{Dirac}\in \Z$. 

Following \cite{CV11} we say that our $\cn=2$ model has the \emph{quiver property} if we may find a set of generators $\{e_v\}$ of $\Gamma$ such that the charge vectors $\gamma\in\Gamma$ of all the BPS particles satisfy
\begin{equation}
 \gamma\in \Gamma_+\quad\text{or}\quad-\gamma\in \Gamma_+,
\end{equation}
where $\Gamma_+\equiv \oplus_v \,\Z_+\,e_v$ is the \emph{positive cone} in $\Gamma$.
Given a $\cn=2$ theory with the quiver property, we associate a $2$--acyclic quiver $Q$ to the data $(\Gamma_+, \langle\cdot,\cdot\rangle_\text{Dirac})$: to each positive generator $e_v$ of $\Gamma_+$ we associate a node $v$ of $Q$ and we 
connect the nodes $u$ $v$ with $B_{uv}$ arrows $u\rightarrow v$ (a negative number meaning arrows in the opposite direction). 
The positive cone $\Gamma_+\subset\Gamma$ is then identified with the cone of dimension vectors of 
 the representations $X$ of $Q$ through $\mathbf{dim}\,X\equiv \sum_v \dim X_v\, e_v$.

The emergence of the quiver $Q$ may be understood as follows. Fix a particle with charge 
$\gamma= \sum_v N_v\,e_v\in\Gamma_+$; on its word--line we have a one dimensional supersymmetric theory 
with 4 supercharges, and the BPS particles correspond to states which are \textsc{susy} vacua of this 1d theory. 
The 1d theory turns out to be a quiver theory in the sense that its K\"ahler target space is the representation 
space of $Q$ of dimension $\sum_v N_v \,e_v$ 
\begin{equation}\label{ppqD}\prod_{\text{arrows}\atop u\rightarrow v} \C^{N_{u} N_{v}}\Big/\Big/ 
\prod_{\text{nodes}\atop v} GL(N_v,\C)\qquad \begin{footnotesize}\text{(symplectic quotient)}\end{footnotesize}.\end{equation}
To completely define the 1d theory we need to specify a $\prod_\text{nodes}  GL(N_v,\C)$--invariant superpotential
$\cw$ (and the FI terms implicit in \eqref{ppqD}); gauge invariance requires $\cw$ to be a function of the traces 
of the products of the bi--fundamental Higgs fields along the closed oriented loops in $Q$. It turns out 
that this function must be linear (a sum of single--trace operators) and thus canonically identified with a linear
 combination (with complex coefficients) of the oriented cycles in $Q$. Thus $\cw$ is a potential for the quiver $Q$ 
in the sense of DWZ \cite{derksen1}.
One shows \cite{ACCERV2} that a 1d configuration is a classical \textsc{susy} vacuum if and only if 
the bi--fundamental Higgs fields associated to the arrows of $Q$ form a \emph{stable} module $X$
 of the Jacobian algebra\footnote{\ A module $X\in\mathsf{mod}\mathscr{J}\!(Q,\cw)$ of dimension $\sum_v N_v e_v$ is
 specified by giving, for each arrow $u \xrightarrow{\ \alpha\ } v$, an $N_v\times N_u$ matrix $X_\alpha$ such that
 the matrices $\{X_\alpha\}$ satisfy the relations $\partial_{X_{\beta}}\cw(X_\alpha)=0$ for all arrows $\beta$ in $Q$.
 Two such representations are \emph{isomorphic}
 if they are related by a $\prod_v GL(N_v,\C)$ transformation.}
\begin{equation}
 \mathscr{J}\!(Q,\cw):=\C Q/(\partial\cw),
\end{equation}
and two field configurations are physically equivalent iff the corresponding modules are isomorphic.
\textit{Stability} is defined in terms of the central charge $Z$ of the $\cn=2$ \textsc{susy} algebra. Being
 conserved, $Z$ is a linear combinations of the various charges; hence may be seen as a linear map
 $Z\colon \Gamma\rightarrow \C$. We assume $\mathrm{Im}\,Z(\Gamma_+) \geq0$, so that we have a well--defined
 function $\arg Z\colon \Gamma_+\rightarrow [0,\pi]$. 
Then $X\in\mathsf{mod} \mathscr{J}\!(Q,\cw)$ is \emph{stable} (with respect to the given central charge $Z$) iff, 
for all proper non--zero submodules $Y$, $\arg Z(Y)<\arg Z(X)$. In particular, \emph{$X$ is stable} $\Rightarrow$
 \emph{$X$ is a brick,} (a module $X$ of an associative algebra is called a \emph{brick} if $\mathrm{End}\,X=\C$).
The isoclasses of stable modules of given dimension $\gamma$ typically form a family parameterized by a K\"ahler
 manifold $\cm_\gamma$; from the viewpoint of the 1d theory the space $\cm_\gamma$ corresponds to zero--modes
 which should be quantized producing $SU(2)_\text{spin}\times SU(2)_R$ quantum numbers. In particular, a $d$--dimensional
 family corresponds (at least) to a BPS supermultiplet with spin content $(0,\tfrac{1}{2})\otimes \frac{d}{2}$ (thus rigid 
modules corresponds to hypermultiplets, $\mathbb{P}^1$--families to vector supermultiplets, and so on).  
Notice that the full dependence of the BPS spectrum from the parameters of the theory is encoded in the central charge $Z$,
 which depends on these parameters as specified by the Seiberg--Witten geometry.
 
For a given $\cn=2$ theory $(Q,\cw)$ is \emph{not} unique; indeed there may be several sets of generators $\{e_v\}$ with 
the above properties. Two allowed $(Q,\cw)$ are related by a Seiberg duality, which precisely coincides with the mutations
 of a quiver with potential in the sense of cluster algebras \cite{derksen1} (this, in particular, requires $\cw$  to be 
non--degenerate in that sense). Therefore, to a QFT we associate a full \textit{mutation class} of quivers. If the mutation class 
is finite we say that the corresponding $\mathcal{N}=2$ QFT is \emph{complete} \cite{CV11} which, in particular, 
implies that no BPS state has spin larger than $1$.
\medskip

\underline{\textbf{$T_2$--duality.}}\label{t2duality} The Seiberg duality/DWZ mutation is not the only source of 
quiver non--uniqueness. The quiver mutations preserve both the number of nodes and $2$--acyclicity. There are more general 
 dualties which do not share these properties. As an example consider the Gaiotto theory corresponding to the $A_1$ $(2,0)$ 6d
theory on a sphere with 3 regular punctures (the $T_2$ theory) \cite{Gaiotto}. $T_2$ consists of 4 free hypermultiplets, carrying
 4 flavor charges, which corresponds to a disconnected quiver with 4 nodes and no arrows. On the other hand, we may associate to
 it a quiver with only \emph{three} nodes, each pair of nodes being connected by a pair of opposite arrows
$\rightleftarrows$
 \cite{ACCERV2}. We refer to the equivalence of the two quivers as `$T_2$--duality'.

\section{The $(Q,\cw)$ class associated to an $\cn=2$ theory}

The BPS states correspond to the stable bricks of the Jacobian algebra. This reduces our problem to a standard problem in Representation Theory \textit{provided} we know which $(Q,\cw)$ mutation class is associated to our $\cn=2$ theory. Determing the mutation class for several interesting gauge theories is the main focus of the present note. 

For $\cn=2$ models having a corner in their parameter space with a weakly coupled Lagrangian description, we have a very physical criterion to check whether a candidate pair $(Q,\cw)$ is correct.
 Simply use the category $\mathsf{mod}\mathscr{J}\!(Q,\cw)$ to compute the would--be BPS spectrum in the limit of vanishing YM coupling $g_\mathrm{YM}\rightarrow 0$ and compare the result with the prediction of perturbation theory. 
The weakly coupled spectrum should consist of
\begin{itemize}
 \item finitely many mutually--local states with bounded masses as $g_\mathrm{YM}\rightarrow 0$:
\begin{enumerate}
 \item vector multiplets making \underline{one copy} of the adjoint representation of the gauge group $G$ (photons and $W$--bosons);
\item hypermultiplets making definite (quaternionic) representations $R_a$ of $G$ (quarks);
\end{enumerate}
\item particles non--local relatively to the $W$--bosons with masses $O(1/g^2_\mathrm{YM})$ (heavy dyons).
\end{itemize}
We ask which pairs $(Q,\cw)$ have such a property (the Ringel property \cite{cattoy}).

\subsection{Magnetic charge and weak coupling regime}\label{Dirrrac}

Consider a quiver $\cn=2$ gauge theory having a weak coupling description with gauge group $G$ (of rank $r$). 
We pick a particular pair $(Q,\cw)$ in the corresponding Seiberg mutation--class which is appropriate for the weak 
coupling regime (along the Coulomb branch).  $\mathsf{mod}\mathscr{J}\!(Q,\cw)$ should contain, in particular, 
one--parameter families of representations corresponding to the massive $W$--boson vector--multiplets 
which are in one--to--one correspondence with the positive roots of $G$. We write $\delta_a$ ($a=1,2,\dots, r$)
 for the charge (\textit{i.e.}\! dimension) vector of the $W$--boson associated to the \emph{simple--root} $\alpha_a$ of $G$.

At a generic point in the Coulomb branch we have an unbroken $U(1)^r$ symmetry.
The $U(1)^r$ electric charges, properly normalized so that they are integral for all states, are given by the fundamental
 coroots\footnote{\ $\mathfrak{h}$ stands for the Cartan subalgebra of the complexified  Lie algebra of the gauge group $G$.} 
$\alpha_a^\vee\in\mathfrak{h}$ ($a=1,2,\dots,r$). The $a$--th electric charge of the $W$--boson 
associated to $b$--th simple root $\alpha_b$ then is
\begin{equation}
q_a=\alpha_a(\alpha^\vee_b)=C_{ab}, \qquad \text{(the Cartan matrix of }G).
\end{equation}
Therefore the vector in $\Gamma\otimes \mathbb{Q}$ corresponding to the $a$--th unit electric charge is
\begin{equation}\mathfrak{q_a}=(C^{-1})_{ab}\,\delta_b.
\end{equation}
Then the magnetic weights (charges) of a representation $X$ are given by
\begin{equation}\label{magneticxxx}
m_a(X)\equiv \langle \dim X, \mathfrak{q}_a\rangle_\text{Dirac}=(C^{-1})_{ab}\,B_{ij}\,(\dim X)_i\,(\delta_b)_j.
\end{equation} 

Dirac quantization requires the $r$ linear forms $m_a(\cdot)$ to be \emph{integral} \cite{cattoy}. This integrality condition is quite a strong constraint on the quiver $Q$, and is our main tool to determine it.

At weak coupling, $g_\mathrm{YM}\rightarrow 0$, the central charge takes the classical form \cite{cattoy}
\begin{equation}\label{zweak}
 Z(X)= -\frac{1}{g^2_\mathrm{YM}}\,\sum_i C_a\,m_a(X)+O(1),
\end{equation}
where $C_a=-i \langle \varphi_a\rangle >0$ in the region of the Coulomb branch covered by the quiver $Q$. 
It is convenient to define the light category, 
$\mathscr{L}(Q,\cw)$, as the subcategory of the modules $X\in\mathsf{mod}\mathscr{J}\!(Q,\cw)$ with
$m_a(X)=0$ for all $a$ such that all their submodules have $m_a(Y)\leq 0$. Comparing with the definition of
 stability in \S.\,\ref{qquiq}, we see that
all BPS states with bounded mass in the limit $g_\mathrm{YM}\rightarrow 0$ correspond to modules
 in $\mathscr{L}(Q,\cw)$, and, in facts, for a $\cn=2$ theory which has a weakly coupled Lagrangian
 description the stable objects of $\mathscr{L}(Q,\cw)$ precisely match the perturbative states. 
They are just the gauge bosons, making one copy of the adjoint of $G$, together with finitely
 many hypermultiplets transforming in definite representations of $G$. 
The detailed structure of $\mathscr{L}(Q,\cw)$ is described in \cite{cattoy}.
\medskip

\textbf{Remarks and Properties}

\begin{enumerate}
 \item $\mathsf{mod}\mathscr{J}\!(Q,\mathcal{W})$ contains \textit{many} ligh subcategories, one for each weakly coupled corner. \textit{E.g.}\! $SU(2)$ $N_f=4$ has a $SL(2,\mathbb{Z})$ orbit of such subcategories;
\item $m(\Gamma_+)\not\geq 0$ $\Rightarrow$ the light category is \emph{not} the restriction to a subquiver, and its quiver is \emph{not} necessarily $2$--acyclic (as in the $T_2$ case \cite{ACCERV2,cattoy}); 
 \item the category $\mathscr{L}(Q,\mathcal{W})$ is \emph{tame} (physically: no light BPS state of spin $>1$);
\item \textit{universality of the SYM sector}: for given gauge group $G$
\begin{equation*}
 \mathscr{L}(Q_\mathrm{SYM},\mathcal{W}_\mathrm{SYM})\subset \mathscr{L}(Q,\mathcal{W})
\end{equation*}
where $(Q_\mathrm{SYM},\mathcal{W}_\mathrm{SYM})$ is the pair for pure $G$ SYM. Only finitely 
many bricks $X\in \mathscr{L}(Q,\mathcal{W})$ and $ X\not\in  \mathscr{L}(Q_\mathrm{SYM},\mathcal{W}_\mathrm{SYM})$, 
they correspond to `quarks'.
\end{enumerate}

\section{First examples}

As a warm--up we consider four classes of (simple) examples.

\subsection{Example 1: $SU(2)$ SQCD with $N_f\leq 4$}

These examples are discussed in detail in \cite{CV11,ACCERV2,cattoy}; here we limit ourselves to a description of the resulting categories.
 One shows \cite{cattoy} that the category $\mathsf{mod}\mathscr{J}\!(Q,\cw)$ is Seiberg--duality equivalent to the Abelian category
 $\mathrm{Coh}(\mathbb{P}^1_{N_f})$ of coherent sheaves on 
$\mathbb{P}^1_{N_f}$ which is $\mathbb{P}^1$ with $N_f$ `double points', that is, the variety in the weighted projective space
$W\mathbb{P}(2,2,\dots,2,1,1)$ of equations
\begin{equation}
 X_i^2-\lambda_i\,X_{N_f+1}-\mu_i\,X_{N_f+2}=0,\qquad i=1,2,\dots, N_f,\quad (\lambda_i:\mu_i)\in\mathbb{P}^1.
\end{equation}
In $\mathrm{Coh}(\mathbb{P}^1_{N_f})$ we have two 
 quantum numbers, \textit{degree} and \textit{rank}
\begin{equation}
\text{rank = magnetic charge,}\qquad \text{degree = $2\times$ electric charge}.
\end{equation}
The light subcategory $\mathrm{Coh}(\mathbb{P}^1_{N_f})\supset\mathscr{L}=\{\text{sheaves of finite length}\}$ \textit{a.k.a.}\! `skyscrapers',
while the dyons correspond to line bundles of various degree.

For $N_f=4$ the curve $\mathbb{P}^1_4$ is Calabi--Yau, hence an elliptic curve $E$. The moduli space of the degree 1 skyscrapers, which is the curve $E$ itself, is isomorphic to its Jacobian $J(E)$ which parameterizes the line bundles of fixed degree. Quantization of $J(E)$ then produces magnetic charged vector--multiplets. Of course,     
 $E\sim J(E)$ reflects the $S$--duality of the theory. See \cite{cattoy} for more details.

\subsection{Example 2: SYM with a simply--laced gauge group $G$}\label{99ert}

The quiver exchange matrix $B$ is fixed by the Dirac charge quantization \cite{cattoy} (cfr.\! \S.\,\ref{Dirrrac}). The standard quiver (the square form) corresponds to
\begin{equation}
 B_\square= C\otimes S, \qquad \text{where }\ \begin{cases}C\ \text{is the Cartan matrix of }G,\\ S\ \text{is the modular $S$--matrix.}\end{cases}
\end{equation}
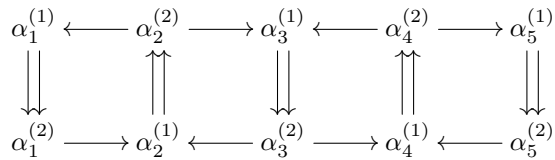
\begin{figure}
\begin{equation*}
\begin{gathered}\xymatrix{\alpha_1^{(1)} \ar@<0.5ex>[d]\ar@<-0.5ex>[d] & \alpha_2^{(2)}\ar[l]\ar[r] & \alpha_3^{(1)} \ar@<0.5ex>[d]\ar@<-0.5ex>[d] & \alpha_4^{(2)}\ar[l]\ar[r] & \alpha_5^{(1)} \ar@<0.5ex>[d]\ar@<-0.5ex>[d]\\
\alpha_1^{(2)}\ar[r] &\alpha_2^{(1)}\ar@<0.5ex>[u]\ar@<-0.5ex>[u] & \alpha_3^{(2)}\ar[l]\ar[r] & \alpha_4^{(1)}\ar@<0.5ex>[u]\ar@<-0.5ex>[u] & \alpha_5^{(2)}\ar[l]}\end{gathered}
\end{equation*}
\caption{\label{yzzrre} The \emph{square} form of the quiver for pure $SU(6)$ SYM}
\end{figure}

The {square} quiver is represented (for $G=SU(6)$) in figure \ref{yzzrre}; it is supplemented by a quartic superpotential $\cw$ \cite{ACCERV2,cattoy}. The charge vector of the $a$--th simple root $W$--boson is equal to $\delta_a\equiv \alpha_a^{(1)}+\alpha_a^{(2)}$, \textit{i.e.}\, the $a$--th simple--root $W$ bosons corresponds to the $\mathbb{P}^1$--family of bricks associated with the 
minimal imaginary root of the $a$--th $\widehat{A}(1,1)$ affine subquiver $\upuparrows_a$. The $a$--th magnetic charge (weight) is (cfr.\! eqn.\eqref{magneticxxx})
\begin{equation}
 m_a(X)= \dim X_{\alpha_a^{(1)}}-\dim X_{\alpha_a^{(2)}}.
\end{equation}
From the discussion around eqn.\eqref{zweak}, the light subcategory $\mathscr{L}^\mathrm{YM}(G)$ containing the perturbative BPS spectrum
 is then given by the modules $X\in\mathsf{mod}\mathscr{J}\!(Q,\cw)$ with $m_a(X)=0$ such that all their submodules $Y$ satisfy $m_a(Y)\leq 0$, $\forall a$.

\begin{figure}
\begin{align*}
& \begin{gathered}
\xymatrix@R=2.0pc@C=3.0pc{
&\ar@(ul,dl)[]_{A_{1}} \alpha_1 \ar@/_0.7pc/[r]_{\widetilde{\psi}_{1}} &\ar@/_0.7pc/[l]_{\psi_{1}} \ar@(dr,dl)[]^{A_2} \alpha_2 \ar@/_0.7pc/[r]_{\widetilde{\psi}_2} &\ar@/_0.7pc/[l]_{\psi_{2}} \ar@(dr,dl)[]^{A_3} \alpha_3 \ar@/_0.7pc/[r]_{\widetilde{\psi}_3}&\ar@/_0.7pc/[l]_{\psi_{3}} \ar@(dr,dl)[]^{A_4} \alpha_4 \ar@/_0.7pc/[r]_{\widetilde{\psi}_4}&\ar@/_0.7pc/[l]_{\psi_{4}} \ar@(ur,dr)[]^{A_5} \alpha_5  
}
\end{gathered}
\end{align*}
\caption{\label{0023c} The reduced quiver $Q^\prime$ for $SU(6)$ pure SYM.}
\end{figure}
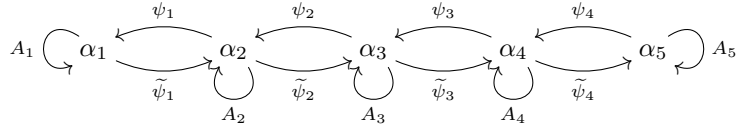

We may break $G\rightarrow SU(2)_a\times U(1)^{r-1}$ at weak coupling and describe the Higgs mechanism perturbatively; that is, the gauge breaking should respect the light \emph{sub}category. Mathematically, this gives the following result at the level of Abelian categories of modules
\begin{equation}
 X\in \mathscr{L}^\mathrm{YM}(G) \ \Rightarrow \ X\big|_{\upuparrows_a}\in \mathscr{L}^\mathrm{YM}(SU(2))\quad\forall\,a,
\end{equation}
which may be checked directly.
Then, if $X$ is indecomposable, in each Kronecker subquiver $\upuparrows_a$ we may set one of the arrows to $1$
with the result that the category $\mathscr{L}^\mathrm{YM}(G)$ gets identified with the category of modules of a Jacobian algebra
\begin{equation}
\mathscr{L}^\mathrm{YM}(G)=\mathsf{mod}\mathscr{J}\!(Q^\prime,\mathcal{W}^\prime)
\end{equation}
where the \emph{reduced} quiver $Q^\prime$ is the \textit{double}\footnote{\ Given an unoriented graph $L$, its
 \textit{double quiver} $\overline{L}$ is obtained by replacing each edge $a$ of $L$ by a pair of opposite 
arrows $\xymatrix{\bullet \ar@<0.4ex>[rr]^{\psi_a}&&\bullet \ar@<0.4ex>[ll]^{\widetilde{\psi}_a}}$. To write eqn.\eqref{eerrq}
 we have picked an arbitrary orientation of $G$, the algebra $\mathscr{J}(Q^\prime,\cw^\prime)$ 
being independent of choices, up to isomorphism.} of the  Dynkin graph\footnote{ By abuse of notation,
we use the same symbol $G$ for the gauge group and its Dynkin graph.} $G$ with loops $A_v$ attached at the nodes (i.e.\! 
the `$\mathcal{N}=2$ quiver' of $G$), see figure \ref{0023c} for the $SU(6)$ example.
 The reduced quiver $Q^\prime$ is equipped with the superpotential
\begin{equation}\label{eerrq}
 \mathcal{W}^\prime=\sum_{a:\;\overrightarrow{\text{edges}}\in G} \mathrm{tr}\big(\widetilde{\psi}_aA_{t(a)}\psi_{a}-{\psi}_aA_{h(a)}\widetilde{\psi}_{a}\big). 
\end{equation}

Given a module $X\in\mathsf{mod}\mathscr{J}\!(Q^\prime,\mathcal{W}^\prime)$, consider the linear map
\begin{equation}
\ell\colon(X_{\alpha_1},X_{\alpha_2},\cdots, X_{\alpha_r})\mapsto (A_1 X_{\alpha_1},A_2 X_{\alpha_2},\cdots, A_r X_{\alpha_r}). 
\end{equation}
It is easy to check that $\ell\in \mathrm{End}\, X$, hence $X$ a brick $\Rightarrow$ $A_i=\lambda\in \mathbb{C}$ for all $i$ (in fact, $\lambda\in \mathbb{P}^1$).
Fixing $\lambda\in \mathbb{P}^1$, the brick $X$ is identified with a brick of the double $\overline{G}$ of the Dynkin graph\footnote{\ The reduced quiver is \underline{not} $2$--acyclic: this is related to the fact that it describes a subset of states which are \textit{all} mutually local, hence have trivial Dirac pairing. At the level of the quiver this means that the \emph{net} number of arrows from node $i$ to node $j$ must vanish (while we need to have arrows since the perturbative sector is not a free theory). }
\begin{align}
\overline{A_5}
\hskip -0.8cm\begin{gathered}
\xymatrix@R=2.0pc@C=3.0pc{
& \alpha_1 \ar@/_0.7pc/[r]_{\widetilde{\psi}_{1}} &\ar@/_0.7pc/[l]_{\psi_{1}}  \alpha_2 \ar@/_0.7pc/[r]_{\widetilde{\psi}_2} &\ar@/_0.7pc/[l]_{\psi_{2}}  \alpha_3 \ar@/_0.7pc/[r]_{\widetilde{\psi}_3}&\ar@/_0.7pc/[l]_{\psi_{3}}  \alpha_4 \ar@/_0.7pc/[r]_{\widetilde{\psi}_4}&\ar@/_0.7pc/[l]_{\psi_{4}}  \alpha_5  
}
\end{gathered}
\end{align}
subjected to relations
\begin{equation}\label{rerrt}
 \sum_{t(a)=v} \psi_a\widetilde{\psi}_a-\sum_{h(a)=v}\widetilde{\psi}_a\psi_a=0.
\end{equation}
The algebra defined by the double quiver $\overline{G}$ with the relations
\eqref{rerrt} is known as the Gelfand--Ponomarev \textit{preprojective algebra} of the graph $G$, written $\mathcal{P}(G)$ \cite{pre1}.
There are three basic results on the preprojective algebra of a graph $L$:
\begin{itemize}
 \item Gelfand and Ponomarev \cite{pre1}: $\dim \mathcal{P}(L)<\infty$ if and only if $L$ is an $ADE$  Dynkin graph;
\item Crawley--Boevey \cite{CBlemma}: Let $C_L=2-I_L$ be the Cartan matrix of the graph $L$. Then for all $X\in \mathsf{mod}\,\mathcal{P}(L)$ 
\begin{equation}\label{xxx4561}
 2\, \dim\mathrm{End}\, X= (\mathbf{dim}\, X)^t\, C_L(\mathbf{dim}\, X)+\dim \mathrm{Ext}^1(X,X)
\end{equation}
\item Lusztig \cite{LU}: Let $X$ be an indecomposable module of $\mathcal{P}(L)$ belonging to a family of non--isomorphic ones parameterized by the (K\"ahler) moduli space $\cm(X)$. Then  \begin{equation}\label{xxx4562}\dim \mathcal{M}(X)=\tfrac{1}{2}\dim \mathrm{Ext}^1(X,X).\end{equation}
\end{itemize}

If $L$ is an $ADE$ graph $G$, the integral quadratic form $v^t\, C_G\, v$ is positive--definite and even; then $X\neq 0$ implies $(\mathbf{dim}\, X)^t\, C_L(\mathbf{dim}\, X)\geq 2$ with equality if and only if $\mathbf{dim}\,X$ is a positive root of $G$.
From eqns.\eqref{xxx4561}\eqref{xxx4562} it follows that if $X$ is a brick of $\mathcal{P}(G)$ it must be rigid with $\mathbf{dim}\,X$ a positive root of $G$.
 Going back to $\mathscr{L}^\mathrm{YM}(G)$, we see that a module in the light category is a brick iff $\mathbf{dim}\, X$ is a positive root of $G$ and 
$\mathcal{M}(X)=\mathbb{P}^1$. By the dictionary between physics and Representation Theory, this means that
 the BPS states which are stable and have bounded mass as $g_\mathrm{YM}\rightarrow 0$ are 
\textit{vector--multiplets in the adjoint of} the gauge group $G$. In fact, a more detailed analysis shows \cite{cattoy} that there is precisely one copy of the adjoint in each weakly coupled BPS chamber. This is, clearly, the result expected for pure SYM at weak coupling;
in particular, is shows that the identification \cite{CNV} of  $(Q,\mathcal{W})$ is correct.

\subsection{Example 3: SQCD with $G$ simply--laced and $N_a$ quarks in the $a$--th fundamental representation}

We consider $\cn=2$ SQCD with a simply--laced gauge group $G=ADE$ coupled to $N_a$ \emph{full} hypermultipletss in the representation $F_a$ with Dynkin label $[0,\cdots, 0, 1,0,\cdots,0]$ ($1$ in the $a$--th position, $a=1,2,\dots, r$).
The prescription for the quiver is simple \cite{ACCERV2}: one replaces the $a$--th Kronecker subquiver $\downdownarrows_a$ of the pure $G$ SYM quiver (cfr.\! \S.\,\ref{99ert}) as follows
\begin{equation}\label{p223r}
\begin{gathered}\xymatrix{\ar@{..}[r] &\bullet\ar@<0.4ex>[dd]
^{B_a}\ar@<-0.4ex>[dd]_{A_a}\ar@{..}[r] &\\
\\
\ar@{..}[r] &\bullet\ar@{..}[r]  & }\end{gathered}\ \
\xrightarrow{\phantom{mmmmmmmm}}\ \ 
\begin{gathered}\xymatrix{\ar@{..}[r] &\bullet\ar@<0.4ex>[dd]^{B_a}\ar@<-0.4ex>[dd]_{A_a}\ar@{..}[r] &\\
&& &\bullet\ar[ull]^{\phi_1}  &\cdots& \bullet\ar[ullll]_{\phi_{N_a}} 
\\
\ar@{..}[r] &\bullet\ar[rru]^{\widetilde{\phi}_1}\ar[rrrru]_{\widetilde{\phi}_
{N_a}}\ar@{..}[r]  & }\end{gathered}
\end{equation}
and replaces the pure SYM superpotential $\cw_\mathrm{SYM}$ with
\begin{gather}\label{p223rs}
\mathcal{W}\longrightarrow \mathcal{W}_\mathrm{SYM}+\sum_{i=1}^{N_i}\mathrm{tr}\big[(\alpha_i\,A_a-\beta_i\,B_a)\phi_i\,\widetilde{\phi}_i\big],\\ (\alpha_i:\beta_i)\equiv \lambda_a\in\mathbb{P}^1\ \text{pairwise distinct}.
\end{gather}
The exchange matrix of the resulting quiver, $B$, has $N_i$ zero eigenvalues corresponding to the $N_a$ flavor charges carried by the quarks. Formally \cite{ACCERV2}, we may extendend this construction to the case in which we have quarks in several distinct fundamental representations, just be applying the substitutions \eqref{p223r}\eqref{p223rs} to all the corresponding Kronecker subquivers of the (square) pure SYM quiver.

\begin{figure} 
\begin{align*}
& \begin{gathered}
\xymatrix@R=2.0pc@C=3.0pc{& & & h \ar@/_0.7pc/[dd]_{{\phi}_{1}} \\
\\
&\ar@(ul,dl)[]_{A_{1}} \alpha_1 \ar@/_0.7pc/[r]_{\widetilde{\psi}_{1}} &\ar@/_0.7pc/[l]_{\psi_{1}} \ar@(dr,dl)[]^{A_2} \alpha_2 \ar@/_0.7pc/[r]_{\widetilde{\psi}_2} &\ar@/_0.7pc/[l]_{\psi_{2}} \ar@(dr,dl)[]^{A_3} \alpha_3\ar@/_0.7pc/[uu]_{\widetilde{\phi}_{1}}  \ar@/_0.7pc/[r]_{\widetilde{\psi}_3}&\ar@/_0.7pc/[l]_{\psi_{3}} \ar@(dr,dl)[]^{A_4} \alpha_4 \ar@/_0.7pc/[r]_{\widetilde{\psi}_4}&\ar@/_0.7pc/[l]_{\psi_{4}} \ar@(ur,dr)[]^{A_5} \alpha_5  
}
\end{gathered}
\end{align*}
\caption{\label{ll451}The \textit{reduced} quiver $\overline{A_5[3,1]}$ for the light category of $G=SU(6)$ SYM coupled to one hypermultiplet $h$ in the $3$--rd fundamental rep.\! (\textit{i.e.}\! the $\mathbf{20}$).}
\end{figure}
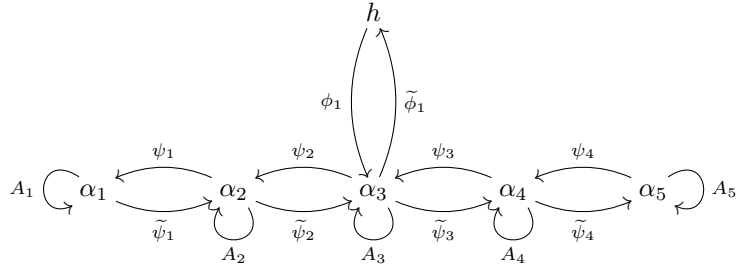

Going through the same steps as in \S.\,\ref{99ert}, one sees that 
the light category $\mathscr{L}=\mathsf{mod}\mathscr{J}\!(Q^\prime, \mathcal{W}^\prime)$ with
 $Q^\prime$ the double of the graph $G[a,N_a]$ obtained by adding $N_a$ extra nodes to the Dynkin graph $G$ connected with a single hedge to the $a$--th node of $G$ and having loops only at all `old' nodes of $G$ \cite{cattoy} (see figure \ref{ll451} for a typical example)
and superpotential
\begin{equation}\mathcal{W}^\prime= \mathcal{W}^\prime_\mathrm{SYM}+\sum\nolimits_i \mathrm{tr}\big[(\alpha_i\,A_a-\beta_i)\phi_i\,\widetilde{\phi}_i\big].\end{equation}

As in \S.\,\ref{99ert}, 
$X$ is a brick $\boldsymbol{\Rightarrow}$ $A_i=\lambda\in\mathbb{P}^1$. Now we have two distinct cases:
\begin{enumerate}
 \item $\lambda$ is \underline{generic} (\textit{i.e.}\! $\lambda\neq \lambda_i$, $i=1,2,\dots, N_a$): the Higgs fields $\phi_i,\widetilde{\phi}_i$ are \textit{massive} and may be integrated out.
Then $X$ is a brick of $\mathcal{P}(G)$ and its charge vector $\mathbf{dim}\,X$ is a positive root of $G$. These are the same representations as for the light category of pure SYM and they correspond to $W$--bosons in the adjoint of $G$;
\item $\lambda=\lambda_a$, then  $X$ is a brick of the preprojective algebra $\mathcal{P}(G[i,1])$. Right properties (finitely many, rigid, in right reprs.\! of $G$) \underline{if and only if} 
 $G[i,1]$ is also a Dynkin graph.
\end{enumerate}
By comparison one gets the following \cite{cattoy}:\medskip

\noindent\textbf{Theorem.} \textit{(1) Consider $\mathcal{N}=2$ SYM with simple simply--laced gauge group $G$ coupled to a hyper in a representation of the form $F_a=[0,\cdots, 0, 1,0,\cdots,0]$. The resulting QFT is \emph{Asymptotically Free} if and only if the augmented graph $G[a,1]$ obtained by adding to the Dynkin graph of $G$ an extra node connected by a single edge to the $a$--th node of $G$ is also an $ADE$ Dynkin graph.
 (2) The model has a Type IIB engineering iff, in addition, the extra node is an extension node in the extended (affine) augmented Dynkin graph $\widehat{G[a,1]}$.}
\medskip

\begin{figure}
\begin{align*}
 &SU(N)\ \text{with }\mathbf{N}  &&\begin{gathered}\xymatrix{\bullet \ar@{-}[r] & \bullet\ar@{-}[r] & \bullet \ar@{-}[r] &\cdots \ar@{-}[r] & \bullet \ar@{..}[r] & 0}\end{gathered}\\
\\
&SU(N)\ \text{with }\mathbf{N(N-1)/2}  &&\begin{gathered} \xymatrix{\bullet \ar@{-}[r] & \bullet\ar@{-}[r] & \bullet \ar@{-}[r] &\cdots \ar@{-}[r] & \bullet \ar@{-}[r] & \bullet\\
& & & & 0\ar@{..}[u]}\end{gathered}\\
&SU(6)\ \text{with }\mathbf{20}  &&\begin{gathered} \xymatrix{\bullet \ar@{-}[r] & \bullet\ar@{-}[r] & \bullet \ar@{-}[r]  & \bullet \ar@{-}[r] & \bullet\\
& &  0\ar@{..}[u]}\end{gathered}\\
&SU(7)\ \text{with }\mathbf{35}  &&\begin{gathered} \xymatrix{\bullet \ar@{-}[r] & \bullet\ar@{-}[r] & \bullet \ar@{-}[r]  & \bullet \ar@{-}[r] & \bullet\ar@{-}[r]& \bullet\\
& &  0\ar@{..}[u]}\end{gathered}\\
&SU(8)\ \text{with }\mathbf{56}  &&\begin{gathered} \xymatrix{\bullet \ar@{-}[r] & \bullet\ar@{-}[r] & \bullet \ar@{-}[r]  & \bullet \ar@{-}[r] & \bullet\ar@{-}[r]& \bullet\ar@{-}[r] &\bullet\\
& &  0\ar@{..}[u]}\end{gathered}\\
&SO(2n)\ \text{with }\mathbf{2n}  &&\begin{gathered}\xymatrix{0 \ar@{..}[r] & \bullet\ar@{-}[r] & \bullet \ar@{-}[r] &\cdots \ar@{-}[r] & \bullet \ar@{-}[r] & \bullet\\
& & & & \bullet\ar@{-}[u]}\end{gathered}\\
&SO(10)\ \text{with }\mathbf{16}  &&\begin{gathered} \xymatrix{0\ar@{..}[r] & \bullet\ar@{-}[r] & \bullet \ar@{-}[r]  & \bullet \ar@{-}[r] & \bullet\\
& &  \bullet\ar@{-}[u]}\end{gathered}\\
&SO(12)\ \text{with }\mathbf{32}  &&\begin{gathered} \xymatrix{0 \ar@{..}[r] & \bullet\ar@{-}[r] & \bullet \ar@{-}[r]  & \bullet \ar@{-}[r] & \bullet\ar@{-}[r]& \bullet\\
& &  \bullet\ar@{-}[u]}\end{gathered}\\
&SO(14)\ \text{with }\mathbf{64}  &&\begin{gathered} \xymatrix{0 \ar@{..}[r] & \bullet\ar@{-}[r] & \bullet \ar@{-}[r]  & \bullet \ar@{-}[r] & \bullet\ar@{-}[r]& \bullet\ar@{-}[r] &\bullet\\
& &  \bullet\ar@{-}[u]}\end{gathered}\\
&E_6\ \text{with }\mathbf{27}  &&\begin{gathered} \xymatrix{\bullet \ar@{-}[r] & \bullet\ar@{-}[r] & \bullet \ar@{-}[r]  & \bullet \ar@{-}[r] & \bullet\ar@{..}[r]& 0\\
& &  \bullet\ar@{-}[u]}\end{gathered}\\
&E_7\ \text{with }\mathbf{56}  &&\begin{gathered} \xymatrix{\bullet \ar@{-}[r] & \bullet\ar@{-}[r] & \bullet \ar@{-}[r]  & \bullet \ar@{-}[r] &\bullet \ar@{-}[r] &\bullet\ar@{..}[r]& 0\\
& &  \bullet\ar@{-}[u]}\end{gathered}\\
\end{align*}
\caption{\label{kk34} The augmented graphs $G[a,1]$ corresponding to  pairs of gauge group $G=ADE$ and fundamental representation which give an asymptotically free $\cn=2$ gauge theory.}
\end{figure}
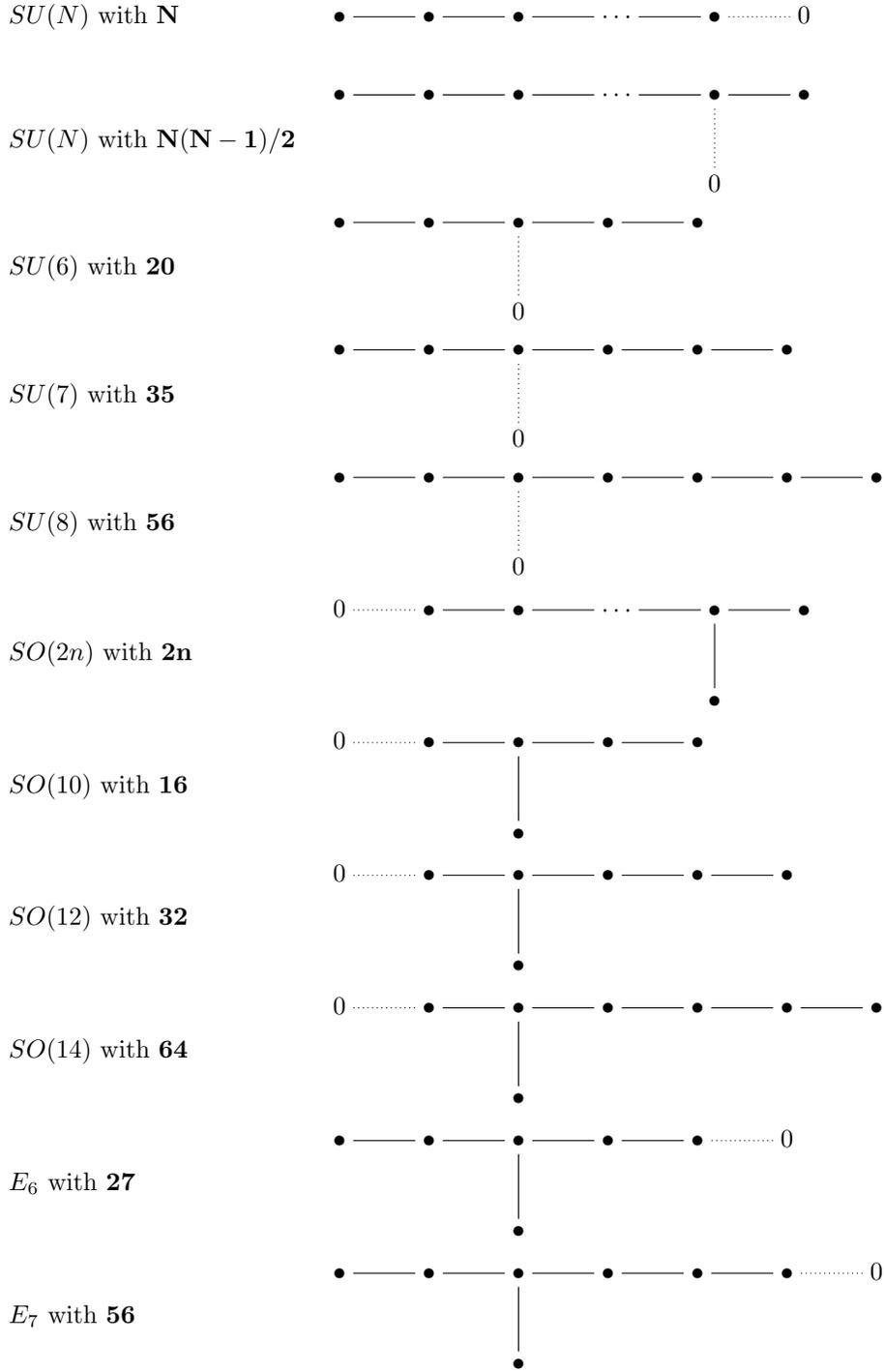

See figure \ref{kk34} for the full list of asymptotically free theories of this class.
Note that in case (2) the light category automatically contains hypermultiplets in the right representation of $G$ since, if $a$ is an extension node in $\widehat{G[a,1]}$ we have
\begin{equation}
 \mathrm{Ad}(G[a,1])=\mathrm{Ad}(G)\oplus [0,\cdots, 0,1,0,\cdots, 0]\oplus \overline{[0,\cdots, 0,1,0,\cdots, 0]}\oplus \text{singlets}.
\end{equation}

 Besides those in figure \ref{kk34} there is another asymptotically free pair (group, representation), 
namely $SU(N)$ with the two--index \emph{symmetric} representation (which is not \emph{fundamental}) 
whose augmented graph is identified with
 the non--simply--laced Dynkin graph of type $B_N$ \cite{nons}.

\subsection{Example 4: $G$ non--simply--laced}

The Dynkin graph of a non--simply laced Lie group $G$ arises by folding a parent  simply--laced Dynkin graph $G_\text{parent}$ along an automorphism group $U$. Specifically, the $G_\text{parent}\rightarrow G$ foldings are
\begin{equation}\begin{aligned}
&D_{n+1}\longrightarrow B_n
&&A_{2n-1}\longrightarrow C_n\\
&D_{4}\longrightarrow G_2
&&E_{6}\longrightarrow F_4.
\end{aligned}\end{equation}
$U=\Z_2$ in all cases except for $D_4\rightarrow G_2$ where it is $\Z_3$. To each node of the folded Dynkin diagrams there is attached an integer $d_a$, namely the number of nodes of the parent graph which were folded into it. This number corresponds to one--half the length--square of the corresponding simple co--root $\alpha_a^\vee$
\begin{equation}\label{defdi}
d_a= \frac{1}{2}(\alpha_a^\vee,\alpha_a^\vee)\equiv \frac{2}{(\alpha_a,\alpha_a)}\qquad a=1,2,\dots, r.
\end{equation}

In general, the light category of a (quiver) $\cn=2$ gauge theory with group $G$ has the structure
\begin{equation}
 \mathscr{L}=\bigvee_{\lambda\in\mathbb{P}^1/U}\mathscr{L}_\lambda
\end{equation}
with $U$ acting on the category $\mathscr{L}_\lambda$ through monodromy functors $\mathscr{M}_u$ \cite{nons}
\begin{equation}\label{yyyq}
 \mathscr{L}_{u\cdot \lambda}= \mathscr{M}_u(\mathscr{L}_\lambda)\qquad u\in U.
\end{equation}
Since the cylinder $\C^*\subset \mathbb{P}^1$ is identified with the Gaiotto plumbing cylinder
 associated to the gauge group $G$, this monodromical construction is equivalent to the geometric
 realization of the non--simply--laced gauge groups in the Gaiotto framework \cite{tackns} or in 
F--theory \cite{BIKMSV}. In the simply--laced case the light category was described in terms of 
the preprojective algebra of $G$; likewise, to each gauge group $G=BCFG$ we may associate a
 generalized `preprojective' algebra of the form
$\mathscr{J}\!(Q^\prime,\cw^\prime)$. $Q^\prime$ is the same reduced quiver as in the $A_r$ 
case (see figure \ref{0023c} for the $r=5$ example) while the reduced superpotential is
\begin{equation}\label{msuper}
\mathcal{W}=\sum_{a\xrightarrow{\alpha}b} \Big( \alpha\, A_{s(\alpha)}^{n(\alpha)}\,\alpha^*-\alpha^*\, 
A_{t(\alpha)}^{m(\alpha)}\,\alpha\Big),
\end{equation} 
where the sum is over the edges $\xymatrix{a \ar@{-}[r]^\alpha&b}$ of $A_r$  and
\begin{equation}
\big(n(\alpha),\: m(\alpha)\big)= \left(\frac{d_a}{(d_a,d_b)},\: \frac{d_b}{(d_a,d_b)}\right).
\end{equation}
One checks \cite{nons} that 
$\mathsf{mod}\mathscr{J}\!(Q^\prime,\cw^\prime)$ has the monodromic property \eqref{yyyq} and 
the dimension vectors of its bricks are the positive roots of $G$, so that the light category corresponds to vector
 multiplets forming a single copy of the adjoint of $G$, as required for pure SYM. 
From the light subcategory $\mathsf{mod}\mathscr{J}\!(Q^\prime,\cw^\prime)$ one reconstructs the full \textit{non--perturbative}
 Abelian category $\mathsf{mod}\mathscr{J}\!(Q,\cw)$, which describes the model in all physical regimes, by using the Dirac integrality conditions described in \S.\,\ref{Dirrrac}. See ref.\!\cite{nons} for details.

\section{Half--hypers}

\subsection{Coupling \emph{full} hypermutliplets to SYM}\label{fullH}

The construction of the pairs $(Q_{N_f},\cw_{N_f})$ for $G=ADE$ SQCD coupled to $N_f$ fundamental \emph{full} hypermultiplets of refs.\!\cite{ACCERV2,cattoy}
was relatively easy: each hypermultiplet has a gauge invariant mass $m_i$, and taking the
  \textit{decoupling limit} $m_i\rightarrow \infty$  we  make $N_f\rightarrow N_f-1$. At the level of modules categories this decoupling processes insets
\begin{gather}\label{reddc}
 \mathsf{mod}\mathscr{J}(Q_{N_f-1},\mathcal{W}_{N_f-1}) \xrightarrow{\subset} \mathsf{mod}\mathscr{J}(Q_{N_f},\mathcal{W}_{N_f})
\end{gather}
as an extension--closed, exact, full, \emph{controlled} Abelian subcategory \cite{cattoy}. In general, a \textit{control function} is a linear map
$\eta\colon \Gamma\rightarrow \Z$, and the controlled subcategory is the full subcategory over the objects $X$ such that $\eta(X)=0$ while for all their subobjects $\eta(Y)\leq 0$. The light subcategory is an example of controlled one with control function the magnetic charge. All decoupling limits of QFT correspond to controlled subcategories in the RT language.

For the decoupling limit $m_i\rightarrow \infty$ the control function
 $f_i\colon \Gamma\rightarrow \mathbb{Z}$ corresponds to the flavor charge dual to $m_i$.
Choosing $f_i$ so that $f_i(\Gamma_+)\geq 0$, we realize $Q_{N_f-1}$ as a full subquiver of $Q_{N_f}$ missing one node, the functor
$\mathsf{mod}\mathscr{J}(Q_{N_f-1},\mathcal{W}_{N_f-1})\rightarrow \mathsf{mod}\mathscr{J}(Q_{N_f},\mathcal{W}_{N_f})$
being the restriction. This gives a recursion relation in $N_f$ of the form
\begin{equation}\label{reddd}
\begin{gathered}\includegraphics{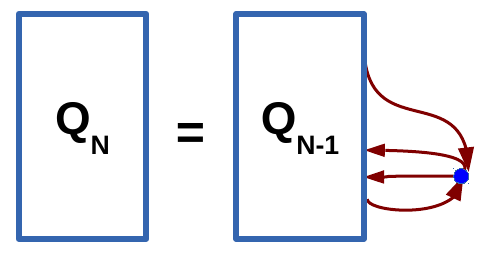}\end{gathered}
\end{equation}
where the blue node in the right corresponds to the controlling flavor charge $f_i$.
By repeated use of this relation, we eventually get to pure $G$ SYM whose quiver is known, see \S.\,\ref{99ert}.
The decoupling process may be easily inverted to get a recursive map $Q_{N_f-1}\rightarrow Q_{N_f}$. 
Indeed, to define such a map we have only to determine the \emph{red} arrows in eqn.\eqref{reddd} which connect $Q_{N_f-1}$ to the
 extra (blue) node in the \textsc{rhs} of \eqref{reddd} which corresponds to an additional massive quark. 
Given the electric weight (\textit{i.e.}\! the $G$--representation) of the added quark, $\omega$,
 the red arrows are uniquely determined by the Dirac pairing of $\omega$ with the charges associated with the nodes of $Q_{N_f-1}$.

This strategy does \underline{not} work for SYM coupled to \emph{half}--hypermultiplets: 
they carry no flavor symmetry, have no mass parameter.
They are
tricky theories, always on the verge of inconsistency:
most of them are indeed quantum inconsistent, but there are a few consistent models which owe their existence to peculiar \textit{`miracles'}. The typical example being $G=E_7$ SYM coupled to half a $\mathbf{56}$.

\subsection{Coupling \emph{half} hypermutliplets}

We use yet another decoupling limit: \textit{extreme Higgs}.
Given a $\mathcal{N}=2$ gauge theory with group $G_r$, of rank $r$, we take a v.e.v.\! 
of the adjoint field $\langle \Phi\rangle\in \mathfrak{h}$ such that
\begin{equation}\label{pp33a}
\alpha_b(\langle \Phi\rangle)=\begin{cases} t\,e^{i\phi},\ \  t\rightarrow+\infty, & b=a\\
O(1) & \text{otherwise}
\end{cases}
\end{equation}
States having electric weight $\rho$ such that $\rho(\langle \Phi\rangle)=O(t)$ decouple, and we remain with a gauge theory with a gauge group $G_{r-1}$ whose Dynkin diagram is obtained by  deleting the $a$--th node from that of $G_r$ 
(coupled to specific matter).
\textit{E.g.}\! starting from $G_7=E_7$ coupled to $\tfrac{1}{2}\,\mathbf{56}$ and choosing $a=1$ we get $G_6=\mathrm{Spin}(12)$ coupled to $\tfrac{1}{2}\,\mathbf{32}$ corresponding 
to deleting the black node in the Dynkin graph
\begin{equation}
\begin{gathered} \xymatrix{\bullet \ar@{-}[r] & \circ\ar@{-}[r] & \circ \ar@{-}[r]  & \circ \ar@{-}[r] &\circ \ar@{-}[r] &\circ\\
& &  \circ\ar@{-}[u]}\end{gathered}
\end{equation}
Again, the decoupling limit should correspond to a controlled Abelian \textit{sub}category of the representations of $(Q_{G_r},\cw_{G_r})$.
One can choose $(Q_{G_r},\mathcal{W}_{G_r})$ in its mutation--class and the phase $\phi$ in \eqref{pp33a} so that the control function $\lambda(\cdot)$ is \emph{non--negative} on the positive--cone $\Gamma_+$.
Then $Q_{G_{r-1}}$
is a full subquiver of $Q_{G_r}$ and $\mathcal{W}_{G_{r-1}}$ is just the restriction of $\mathcal{W}_{G_r}$.
It is easy to see that the complementary full subquiver is a two--nodes Kronecker one $\upuparrows$ \cite{half}. Putting everything together, we get a recursion of the quiver with respect to the rank $r$ of $G_r$  of the form
\begin{equation}
 \begin{gathered}\includegraphics{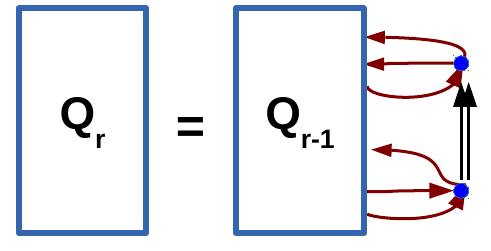}\end{gathered}\end{equation}

If we know the simpler quiver $Q_{G_{r-1}}$, to get $Q_{G_r}$ we need just the fix the red arrows connecting the Kronecker to $Q_{G_{r-1}}$ in the above figure.
Just as in \S.\,\ref{fullH},
the {red} arrows are uniquely fixed by Dirac charge quantization.
Indeed, by the recursion assumption, we know the representations $X_{\alpha_a}$ associated to all simple--root $W$--bosons of $G_r$; under the maximal torus $U(1)^r\subset G$ the simple--root $W$--bosons have charges
$q_a(X_{\alpha_b})=C_{ab}$ (Cartan matrix),
while the dual magnetic charges are given by eqn.\eqref{magneticxxx} which explicitly depends on the red arrows.
It turns out \cite{half} that $m_a(X)\in {}^L\Gamma_\mathrm{root}$ for all $X$  for a \emph{unique} choice of the arrows which are then fixed.
Then $Q_{G_r}$ is uniquely determined if we know $Q_{G_{r-1}}$. $\mathcal{W}_{G_r}$ is also essentially determined, up to some higher--order ambiguity \cite{half}.

\begin{figure}
\begin{equation*}
\xymatrix{ \text{\fbox{\;$E_7\phantom{\Big|}$ $\tfrac{1}{2}\,\mathbf{56}$\;}} & \text{\begin{small}$SU(2)\times SO(12)$\:  $\tfrac{1}{2}(\mathbf{2},\mathbf{12})$\end{small}}\\
 \text{\fbox{\;$SO(12)\phantom{\Big|}$ $\tfrac{1}{2}\,\mathbf{32}$\;}}\ar@{=>}[u] & 
\text{\begin{small}$SU(2)\times SO(10)$\:  $\tfrac{1}{2}(\mathbf{2},\mathbf{10})$\end{small}}\ar@{..>}[ul]\ar@{..>}[u]\\
\text{\fbox{\;$SU(6)\phantom{\Big|}$ $\tfrac{1}{2}\,\mathbf{20}$\;}}\ar@{=>}[u] &
\text{\fbox{\;$SU(2)\times SO(8)\phantom{\Big|}$ $\tfrac{1}{2}\,(\mathbf{2},\mathbf{8})$\;}}\ar@{=>}[ul]\ar@{..>}[u]\\
\text{\fbox{\;$SU(2)\times SU(4)\phantom{\Big|}$ $\tfrac{1}{2}\,(\mathbf{2},\mathbf{6})$\;}}\ar@{=>}[u]\ar@{=>}[ur]\\
\text{\fbox{\;$SU(2)^3\phantom{\Big|}$ $\tfrac{1}{2}\,(\mathbf{2},\mathbf{2},\mathbf{2})$\;}}\ar@{=>}[u]
}
\end{equation*}\caption{\label{ppq}The Higgs breaking chain for various SYM models coupled to half hypers.}\end{figure}
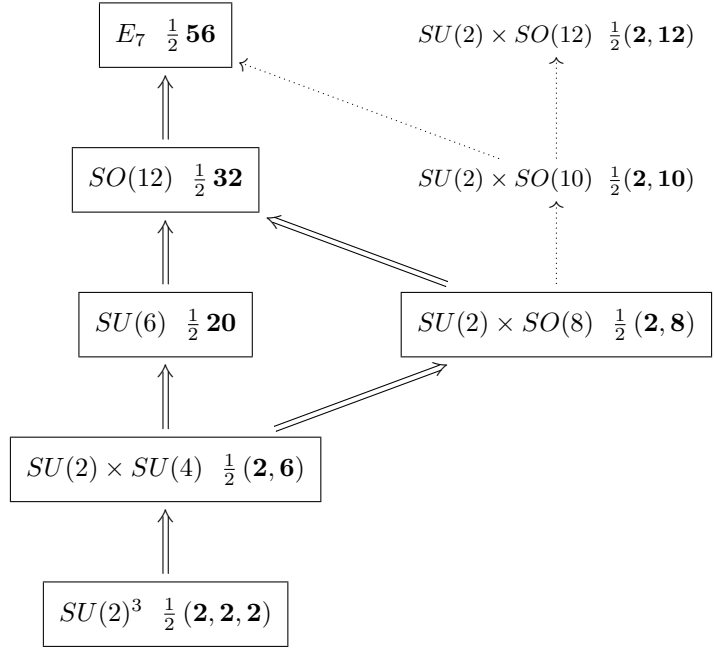

Taking a suitable chain of such Higgs decouplings/symmetry breakings \begin{equation}\label{44rr}G_r\rightarrow G_{r-1}\rightarrow G_{r-2}\rightarrow \cdots\cdots \rightarrow G_k,\end{equation}
we eventually end up with a \emph{complete} $\mathcal{N}=2$ with gauge group $G_k=SU(2)^k$.
The complete $\mathcal{N}=2$ quivers are known by classification \cite{CV11}.
Inverting the Higgs procedure, we may construct the pair $(Q_{G_r},\mathcal{W}_{G_r})$ for the theory of interest by {`pulling back'} through the chain \eqref{44rr} the pair $(Q_\text{max comp},\mathcal{W}_\text{max comp})$ of their {maximal complete} subsector.
For the models of interest the \textit{`pull back'} chain is presented in figure \ref{ppq}. The bottom model $SU(2)^3$ with $\tfrac{1}{2}(\mathbf{2},\mathbf{2},\mathbf{2})$ is complete \cite{CV11,cattoy}.

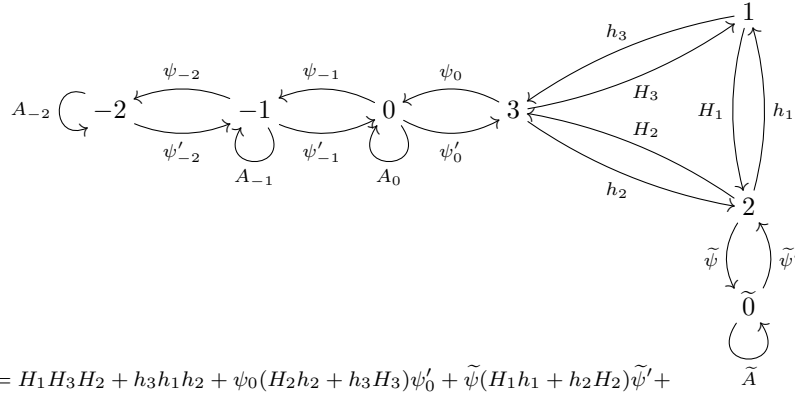
\begin{figure}
\fbox{ $E_7$ $\phantom{\Big|}$coupled to $\tfrac{1}{2}\,\textbf{56}$ }

\begin{equation*}
\begin{gathered}
\xymatrix{
&&&*++[o][F-]{1} \ar[ld]_{\phi} \ar[rr]^{H_1} && *++[o][F-]{2}\ar[dl]_{H_2}\ar[dr]^{\widetilde{\psi}}&\\
*++[o][F-]{\tau_{-2}}\ar@<0.5ex>@{->}[ddd]^{A_{-2}}\ar@<-0.5ex>@{->}[ddd]_{B_{-2}}& *++[o][F-]{\omega_{-1}}\ar[l]_{\psi_{-2}} \ar[r]^{\psi_{-1}^{\prime}}&*++[o][F-]{\tau_0} \ar@<0.5ex>@{->}[ddd]^{A_0}\ar@<-0.5ex>@{->}[ddd]_{B_0} \ar[ddr]^{\phi^{\prime}}& & *++[o][F-]{3}\ar[lu]_{H_3} \ar[ll]_{\psi_0 \qquad} \ar[rr]^{\quad\widetilde{\phi}}&&*++[o][F-]{\widetilde{\tau}}\ar@<0.5ex>@{->}[ddd]^{\widetilde{A}}\ar@<-0.5ex>@{->}[ddd]_{\widetilde{B}}\ar[dddll]_{\widetilde{\phi}^{\prime}}\\
&&&&& &  &&\\
 &&&*++[o][F-]{5}\ar[uuu]_{V_1}\ar[dr]^{h_3} & & *++[o][F-]{6}\ar[ll]_{h_1\qquad}\ar[uuu]_{V_2}&\\
*++[o][F-]{\omega_{-2}}\ar[r]^{\psi_{-2}^{\prime}}&*++[o][F-]{\tau_{-1}} \ar@<0.5ex>@{->}[uuu]^{A_{-1}}\ar@<-0.5ex>@{->}[uuu]_{B_{-1}}&*++[o][F-]{\omega_0}\ar[l]_{\psi_{-1}}  \ar[rr]^{\psi_0^{\prime}}& & *++[o][F-]{4}\ar[ur]^{h_2}\ar[uuu]_{V_3} &&*++[o][F-]{\widetilde{\omega}}\ar[ul]_{\widetilde{\psi}^{\prime}}\\
}
\end{gathered}
\end{equation*}\begin{footnotesize}
\begin{multline*}
 \mathcal{W}_{E_7}= H_1 H_3 H_2 + h_3 h_1 h_2 
+ A \psi V_3 \psi^{\prime} + B \psi H_2 V_2 h_2 \psi^{\prime} 
 + \phi V_1 \phi^{\prime} + \psi V_3 h_3 \phi^{\prime}  +\\ \phi H_3 V_3 \psi^{\prime} B+ \widetilde{A} \widetilde{\psi} V_2 \widetilde{\psi}^{\prime} + \widetilde{B} \widetilde{\psi} H_1 V_1 h_1 \widetilde{\psi}^{\prime}
  + \widetilde{\phi} V_3 \widetilde{\phi}^{\prime} + \widetilde{\psi} V_2 h_2 \widetilde{\phi}^{\prime}  + \widetilde{\phi} H_2 V_2 \widetilde{\psi}^{\prime} \widetilde{B}+\\ +A_0\psi^\prime_{-1}B_{-1}\psi_{-1} -B_0\psi^\prime_{-1}A_{-1}\psi_{-1}+A_{-1}\psi^\prime_{-2}B_{-2}\psi_{-2}-B_{-1}\psi^\prime_{-2}A_{-2}\psi_{-2}
\end{multline*}
\end{footnotesize}
\caption{\label{rrrr4} Quiver and superpotential for the $\cn=2$  $E_7$ SYM coupled to $\tfrac{1}{2}\mathbf{56}$ quark.}
\end{figure}

The pair $(Q_{E_7},\cw_{E_7})$ for the model $G=E_7$ coupled to $\tfrac{1}{2}\,\textbf{56}$ is given in figure \ref{rrrr4}; the other models in figure \ref{ppq} correspond to the restriction to suitable subquivers of $(Q_{E_7},\cw_{E_7})$ \cite{half}. The light category deduced from these pairs contains  light vectors forming one copy of the adjoint of $G$ plus light hypermultiplets in the  $G$--representation $\tfrac{1}{2}\,\mathbf{R}$, with $\mathbf{R}$ irreducible \textit{quaternionic} \cite{half}. Indeed, 
the light category has again the form $\mathsf{mod}\mathscr{J}(Q^\prime,\cw^\prime)$ for a reduced pair $(Q^\prime,\cw^\prime)$. See figure \ref{kkaq} for the the reduced pair for $G=E_7$ coupled to
$\tfrac{1}{2}\mathbf{56}$; the other models are obtained by restriction of this one. Note that $Q^\prime_{E_7}$ (and hence all reduced quivers $Q^\prime_{G_r}$ in the Higgs chain) contains as a full subquiver the quiver of the Gaiotto $A_1$ theory on $S^2$ with 3 punctures (the $T_2$ theory) described in \cite{ACCERV2}. Hence for all these models the `$T_2$--duality' of \S.\,\ref{qquiq} is operative; this duality is crucial --- together with special properties of the relevant Dynkin graphs --- to check the above claims on the BPS spectrum at weak coupling. Details may be found in \cite{half}.

\begin{figure}
\begin{equation*}
\begin{gathered}
\xymatrix@R=2.0pc@C=3.0pc{
&&&&&&1 \ar@/_0.5pc/[dll]_{h_3} \ar@/_0.5pc/[dd]_{H_1}\\
&\ar@(ul,dl)[]_{A_{-2}} -2 \ar@/_0.7pc/[r]_{\psi_{-2}^{\prime}} &\ar@/_0.7pc/[l]_{\psi_{-2}} \ar@(dr,dl)[]^{A_{-1}} -1\ar@/_0.7pc/[r]_{\psi_{-1}^{\prime}}&\ar@/_0.7pc/[l]_{\psi_{-1}} \ar@(dr,dl)[]^{A_0} 0 \ar@/_0.7pc/[r]_{\psi_0^{\prime}} & 3 \ar@/_0.7pc/[l]_{\psi_0} \ar@/_0.5pc/[urr]_{H_3} \ar@/_0.5pc/[drr]_{h_2}&&\\
&&&&&&2 \ar@/_0.5pc/[ull]_{H_2} \ar@/_0.5pc/[uu]_{h_1}\ar@/_0.7pc/_{\widetilde{\psi}}[d]\\
&&&&&&\ar@(ld,rd)[]_{\widetilde{A}}\widetilde 0 \ar@/_0.7pc/_{\widetilde{\psi}^\prime}[u]
}
\end{gathered}
\end{equation*}\vskip-24pt\begin{footnotesize}
\begin{multline*}
 \mathcal{W}^\prime_{E_7}= H_1H_3H_2+h_3h_1h_2+\psi_0(H_2h_2+h_3H_3)\psi_0^\prime+ \widetilde{\psi}(H_1h_1+h_2H_2)\widetilde{\psi}^\prime+\\
+A_0\psi_0\psi_0^\prime+A_0\psi^\prime_{-1}\psi_{-1}-A_{-1}\psi_{-1}\psi^\prime_{-1}+A_{-1}\psi^\prime_{-2}\psi_{-2}-A_{-2}\psi_{-2}\psi_{-2}^\prime
+\widetilde{A}\widetilde{\psi}\widetilde{\psi}^\prime.
\end{multline*}\end{footnotesize}
\caption{\label{kkaq} Reduced pair of the light category $\mathscr{L}_{E_7}\equiv\mathsf{mod}\mathscr{J}(Q^\prime,\cw^\prime)$ for
  $E_7$ SYM with $\tfrac{1}{2}\mathbf{56}$ quark.}
\end{figure}

\end{document}